\documentclass[twocolumn]{article}

\usepackage{microtype} 

\usepackage{footnote}
\makesavenoteenv{tabular}

\usepackage[left=25mm,right=25mm,top=32mm,columnsep=28pt]{geometry}
\usepackage[hang, small,labelfont=bf,up,textfont=it,up]{caption} 
\usepackage{booktabs} 
\usepackage{mathtools}
\usepackage{hyperref}

\usepackage{enumitem} 
\setlist[itemize]{noitemsep} 

\usepackage{abstract}

\usepackage{titlesec} 
\renewcommand\thesection{\Roman{section}} 
\renewcommand\thesubsection{\roman{subsection}}
\titleformat{\section}[block]{\large\scshape\centering}{\thesection.}{1em}{}
\titleformat{\subsection}[block]{\large}{\thesubsection.}{1em}{} 

\usepackage{fancyhdr} 
\usepackage{titling} 

\usepackage{hyperref} 
\usepackage{amsmath}
\newcommand{\norm}[1]{\left\lVert #1 \right\rVert}

\usepackage{graphics}
\usepackage{graphicx}

\usepackage{float}
\usepackage{tablefootnote}
\usepackage{tikz}
\usetikzlibrary{positioning}
\usepackage{adjustbox}
\usepackage{stfloats}
\everymath{\displaystyle}

\usepackage{appendix}

\DeclareMathOperator{\sech}{sech}

\title{\huge\bfseries Using Restricted Boltzmann Machines to Model Molecular Geometries} % Article title
\author{Peter Nekrasov, Jessica Freeze, Victor Batista \\[1ex] % Your name
\normalsize Yale University, Department of Chemistry \\ % Your institution
\normalsize \href{mailto:}{peter.nekrasov@yale.edu}}
\date{\today}

\begin{document}
\maketitle

\begin{abstract}

Precise physical descriptions of molecules can be obtained by solving the Schrodinger equation; however, these calculations are intractable and even approximations can be cumbersome. Force fields, which estimate interatomic potentials based on empirical data, are also time-consuming. This paper proposes a new methodology for modeling a set of physical parameters by taking advantage of the restricted Boltzmann machine's fast learning capacity and representational power. By training the machine on ab initio data, we can predict new data in the distribution of molecular configurations matching the ab initio distribution. In this paper we introduce a new RBM based on the Tanh activation function, and conduct a comparison of RBMs with different activation functions, including sigmoid, Gaussian, and (Leaky) ReLU. Finally we demonstrate the ability of Gaussian RBMs to model small molecules such as water and ethane.

%Here we show that approximations of the electronic ground state are sufficient to describe rudimentary geometric properties of polyatomic molecules by using the water molecule as an

\end{abstract}

%----------------------------------------------------------------------------
%	ARTICLE CONTENTS
%----------------------------------------------------------------------------

\section{Introduction}

Recent innovations in data science have led to a proliferation of machine learning techniques, many of which are used to find patterns and categorize data. Among these models, the restricted Boltzmann machine (RBM) has shown special promise for its ability to quickly learn the probability distribution of a training set and extract its key ``features." The RBM is a versatile tool used in many practical applications ranging from social media services to product recommendations. While RBMs are becoming increasingly popular for conventional problems in data science, they are underutilized within the field of chemistry. Though recently a few sporadic studies have used RBMs to model self-avoiding walks of polymers \cite{yu} and perform quantum electronic structure calculations \cite{xia}, no systematic approach has been taken to develop this software for a multitude of continuous systems. 

There are several features which make the use of RBMs conducive to the field of chemistry. For one, the restricted Boltzmann machine falls under a family of energy-based models, which means it associates an energy value to any given state of the machine. Because chemistry calculations constantly utilize energy terms, the RBM can be adapted to complex physical computations. In this sense, a trained RBM can serve as a Hamiltonian for a physical system \cite{xia}. Furthermore, the RBM's strength lies in its simplicity: with a simple structure and sampling algorithm, the RBM is straightforward to use and efficient to implement.

\begin{figure}[h]
    \centering
    \includegraphics[scale=0.5]{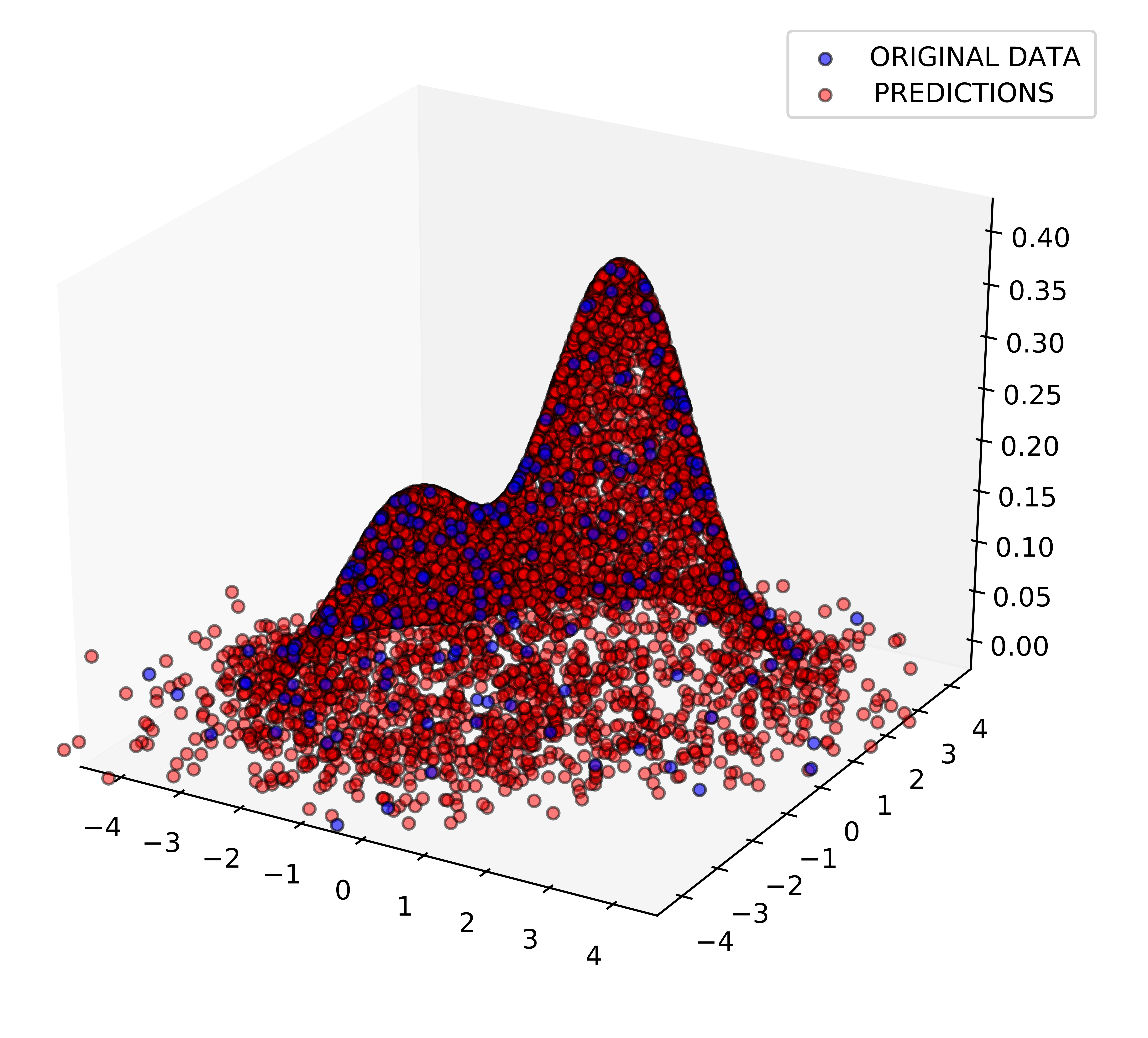}
    \caption{Given a sparse set of data points (blue), our aim is to train the model to predict the shape of the overall distribution and extract new data points (red) which could have been part of the original distribution.}
    \label{fig:sparsity}
\end{figure}

Imagine the following predicament: a set of molecular geometries or quantum states is generated either through experimental data or complex  ab initio calculations. While this dataset is thought to be an accurate representation of the overall distribution, the number of samples is insufficient to perform any meaningful analysis on the system. Furthermore, one would like to extend the given representation to a complete ensemble within the phase space. In this case, RBMs can enable us to learn the overall distribution of physical parameters and predict points not included in the the given dataset (Figure \ref{fig:sparsity}). 

The advantage of this approach is that it requires no prior information about the system in question. Most approximations of the quantum wave function require an expression of the constituent energies of the system. Likewise, molecular dynamics force fields require an understanding of interatomic potentials such as electrostatic and van der Waals forces. Instead, the RBM simply utilizes the statistical frequency of the training configurations to learn an internal representation of their energies. While quantum calculations are laborious to run, the RBM adopts a simple training and sampling algorithm, providing us with resounding representational power at a low computational cost. 

%Instead of computing individual potentials for a multi-particle system, we can use the RBM to learn the energies of different configurations based on their statistical frequency. 

In this paper we describe different types of RBMs and criteria for assessing their performance. We then show how the Gaussian RBM (GRBM) learns diverse sets of training data and reproduces various distributions. Finally, we use GRBMs to represent molecular systems with multiple degrees of freedom to show how they can learn bond and angle energies of small molecules such as  H\textsubscript{2}O and ethane.

%------------------------------------------------

\section{Methods}

\paragraph{Model overview} 

\begin{figure}[h]
    \centering
    \begin{tikzpicture}[shorten >=1pt,->,draw=black!50, node distance=\layersep]
        \tikzstyle{every pin edge}=[<-,shorten <=1pt]
        \tikzstyle{neuron}=[circle,fill=black!25,minimum size=14pt,inner sep=0pt]
        \tikzstyle{input neuron}=[neuron, fill=red!50];
        \tikzstyle{hidden neuron}=[neuron, fill=blue!50];
        \tikzstyle{annot} = [text width=4em, text centered]
    
        % Draw the input layer nodes
        \foreach \name / \y in {1,...,3}
        % This is the same as writing \foreach \name / \y in {1/1,2/2,3/3,4/4}
            \node[input neuron] (I-\name) at (0,-\y - 1.15) {};
    
        % Draw the hidden layer nodes
        \foreach \name / \y in {1,...,8}
            \node[hidden neuron] (H-\name) at (2.0,-\y * 0.7) {};
    
        % Connect every node in the input layer with every node in the
        % hidden layer.
        \foreach \source in {1,...,3}
            \foreach \dest in {1,...,8}
                \draw [<->] (I-\source) -- (H-\dest);

        % Annotate the layers
        \node[annot,above of=H-1, node distance=1cm] (hl) {Hidden layer};
        \node[annot,above of=I-1, node distance=1cm] (hl) {Visible layer};
    \end{tikzpicture} 
    \caption{A sample RBM with three visible nodes and eight hidden nodes (3-8-RBM). Every visible node is connected to every hidden node, and each node has its own bias (not shown).}
    \label{fig:RBM}
\end{figure}
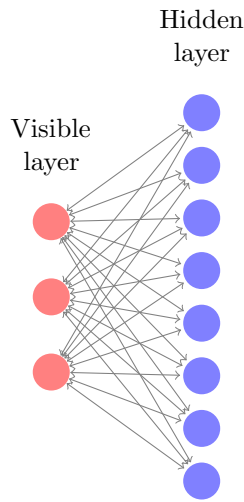

The RBM consists of two layers of neurons, a visible layer and a hidden layer (shown in Figure \ref{fig:RBM}). Every visible node $v_i$ is connected to every hidden node $h_j$ by a set of weights, and each node has its own offset, or bias. The state or value of a given node is dependent on the state of the nodes it is connected to, as well as its bias. For ease of computation, the values of the weights are stored in a weight matrix $W$, where $W_{ij}$ represents the connection from $v_i$ to $h_j$. Meanwhile, the values of the visible and hidden biases are stored in bias vectors $a$ and $b$, respectively. The RBM is ``restricted" in the sense that there are no connections between nodes in the same layer, which simplifies learning \cite{smolensky}.

The visible layer serves as an input to the machine, where the number of visible nodes corresponds to the number of variables that make up the data. Values for the hidden nodes are then calculated by multiplying the visible nodes by the weights, adding the biases, and then applying some sort of activation function. This inference step can be viewed as a transformation from the space of observable parameters to the space represented by the hidden nodes. The method used for calculating between layers is formalized by a set of conditional probabilities that appear in the section below. In this paper we refer to a $n$-$k$-RBM as an RBM with $n$ visible nodes and $k$ hidden nodes. 

%\[ V \xrightarrow{W} H \xrightarrow{W^T} V' \]

A notable feature of the RBM, in contrast with other machine learning models, is that it does not have an ``output" in the normal sense. Depending on the situation, the output of an RBM can be the visible layer, hidden filters, or energy. RBMs are trained in an unsupervised fashion (without labelling the data), so the hidden nodes identify their own labels during the course of training \cite{hinton}. In a famous application of RBMs by Netflix to provide movie recommendations (see the Netflix prize \cite{salakhutdinov}), the RBM was trained on a large set of movie ratings obtained from individual users, where the value of each visible node corresponded to the rating of a given movie. Based on the simple pattern that users who like movies from a certain director or genre are likely to enjoy other movies from that same category, the RBM came to associate hidden nodes with movie genres or directors. In this way, each hidden node elucidates a connection or correlation between visible nodes, which is true for continuous data as well. Note that understanding what each hidden node represents requires additional post hoc analysis. 

\paragraph{Energy based models} 

As an energy based model, the RBM associates a scalar measure of energy to each state of the machine. Usually the energy equation is a combination of each layer times its respective biases and a term which relates the two layers. The goal of training is to minimize the overall energy of the RBM with respect to the training data. Once trained, the RBM associates lower energies with inputs that fall within the training distribution and higher energies with those that fall outside the training distribution. This becomes useful when generating new configurations because the RBM estimates the energy of a proposed configuration based on where it fits with the training distribution.

\begin{figure*}[t]
    \centering
    \begin{tikzpicture}[]
    \node [rectangle, rounded corners, draw=black, minimum width=2.8cm, minimum height=0.7cm] (hello) at (1.25, 0.35) {};
    \node [rectangle, rounded corners, draw=black, minimum width=4cm, minimum height=0.7cm, above right=1.5cm and -1.0cm of hello] (hi) {};
    \node [rectangle, rounded corners, draw=black, minimum width=2.8cm, minimum height=0.7cm, below right=1.5cm and -1.0cm of hi] (ya) {};  
    \node [rectangle, rounded corners, draw=black, minimum width=4cm, minimum height=0.7cm, above right=1.5cm and -1.0cm of ya] (yum) {};
    \node (cond1) at (1.3, 1.3) {$p(h|v)$};
    \node (cond2) at (4.3, 1.3) {$p(v|h)$};
    \node (cond1) at (7.8, 1.3) {$p(h|v)$};
    \node (vis) at (-1.3, 0.35) {Visible layer};
    \node (hid) at (0.5, 2.55) {Hidden layer};
    \node (hid) at (1.25, -0.4) {Initial values};
    \node (hid) at (6.2, -0.4) {Reconstruction};
    \node () at (11.3, 0.35) {\Huge ...};
    \draw[-latex] (hello.north) -- (hi.south);
    \draw[-latex] (hi.south) -- (ya.north);
    \draw[-latex] (ya.north) -- (yum.south);
    \draw[-latex] (yum.south) -- (11, 0.58);
    \draw[red!50,fill=red!50, text=black] (0.4,0.35) circle (0.2) node {$v_1$};
    \draw[red!50,fill=red!50, text=black] (1.0,0.35) circle (0.2) node {$v_2$};
    \draw[red!50,fill=red!50, text=black] (1.6,0.35) circle (0.2) node {$v_3$};
    \draw[red!50,fill=red!50, text=black] (2.2,0.35) circle (0.2) node {$v_4$};
    \draw[blue!35,fill=blue!35,text=black] (2.2,2.55) circle (0.2) node {$h_1$};
    \draw[blue!35,fill=blue!35,text=black] (2.8,2.55) circle (0.2) node {$h_2$};
    \draw[blue!35,fill=blue!35,text=black] (3.4,2.55) circle (0.2) node {$h_3$};
    \draw[blue!35,fill=blue!35,text=black] (4.0,2.55) circle (0.2) node {$h_4$};
    \draw[blue!35,fill=blue!35,text=black] (4.6,2.55) circle (0.2) node {$h_5$};
    \draw[blue!35,fill=blue!35,text=black] (5.2,2.55) circle (0.2) node {$h_6$};
    \draw[red!50,fill=red!50, text=black] (5.2,0.35) circle (0.2) node {$v_1$};
    \draw[red!50,fill=red!50, text=black] (5.8,0.35) circle (0.2) node {$v_2$};
    \draw[red!50,fill=red!50, text=black] (6.4,0.35) circle (0.2) node {$v_3$};
    \draw[red!50,fill=red!50, text=black] (7.0,0.35) circle (0.2) node {$v_4$};
    \draw[blue!35,fill=blue!35,text=black] (7,2.55) circle (0.2) node {$h_1$};
    \draw[blue!35,fill=blue!35,text=black] (7.6,2.55) circle (0.2) node {$h_2$};
    \draw[blue!35,fill=blue!35,text=black] (8.2,2.55) circle (0.2) node {$h_3$};
    \draw[blue!35,fill=blue!35,text=black] (8.8,2.55) circle (0.2) node {$h_4$};
    \draw[blue!35,fill=blue!35,text=black] (9.4,2.55) circle (0.2) node {$h_5$};
    \draw[blue!35,fill=blue!35,text=black] (10,2.55) circle (0.2) node {$h_6$};
    \end{tikzpicture} 
    \caption{The sampling algorithm for a 4-6-RBM consists in using conditional probabilities to alternate between layers. This technique, known as Gibbs sampling, is relevant to many aspects of using RBMs, including estimating the gradient and calculating the error. This RBM happens to consist of four visible nodes and six hidden nodes.}
    \label{fig:sampling}
\end{figure*}
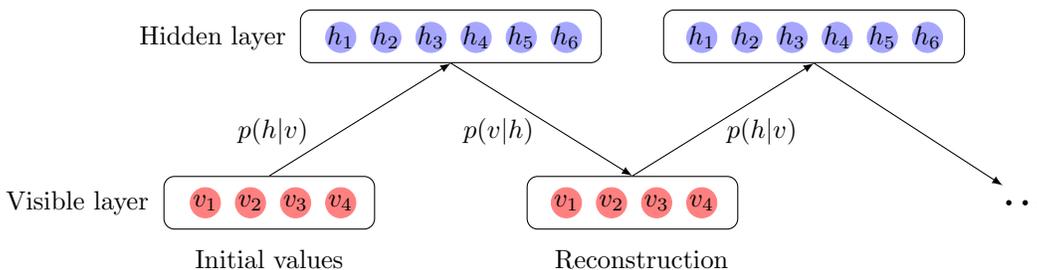

\paragraph{Binary RBM} 

The simplest and most widespread version of the RBM is the binary-binary RBM (BBRBM), which has binary visible and binary hidden units. In the binary-binary RBM, an active unit is represented as 1 while inactive units are represented with 0. For a BBRBM, the energy is given by:
\[ E(v,h) = -a^Tv - b^Th - v^TWh \]
where $v$ is a vector of visible states, $h$ is the hidden states, $a$ is the visible biases, $b$ is the hidden biases, and $W$ is the weight matrix representing the connections between visible and hidden nodes. The visible states $v$ serve as an input to the RBM whereas $a$, $b$, and $W$ are all parameters learned during training.

The joint probability of states $v$ and $h$ is taken from the Boltzmann distribution:
\[p(v,h) = \frac{1}{Z} e^{-E(v,h)} \]
where $Z$ is the partition function, defined as:
\[ Z = \sum_v \sum_h e^{-E(v,h)} \]
which is a sum of over all possible states of the machine. The partition function $Z$ serves to normalize the joint probability distribution so that probabilities sum to 1. From this, we can deduce the marginal probability of a visible configuration $v$, given by the sum of probabilities over all possible hidden configurations:
\begin{align}
    p(v) &= \sum_h p(v,h) \\
    p(v) &= \frac{1}{Z} \sum_h e^{-E(v,h)}
\end{align}
Because the state of the hidden layer depends on the state of the visible layer, we must use a conditional probability $p(h | v)$ to calculate the hidden states. This conditional probability is derived directly from the energy equation \cite{wang}, giving us the activation of hidden states: 
\[ p(h=1|v) = \sigma(b + W^Tv) \]
where $\sigma(x)$ represents the sigmoid activation function. Performing this operation only gives a set of probabilities that each hidden neuron is active; one must then sample a Bernoulli distribution with the given probabilities to reach the actual binary states of the hidden layer. Then, the visible states are computed again from the hidden states using:
\[ p(v=1|h) = \sigma(a + Wh) \]
which gives the probabilities of the visible neurons adopting a value of one. Figure \ref{fig:sampling} shows how conditional probabilities are used to sample back and forth between layers.

A major drawback of the binary RBM is that it is only able to represent binary states or bit strings. As most empirical data is real-valued, the Gaussian RBM was developed to model continuous variables.

\paragraph{Gaussian RBM} The Gaussian RBM (GRBM), also called the Gaussian-binary RBM, is an effective tool for modeling real-valued data. While the hidden layer remains binary, the visible layer can now adopt real values, with an additional parameter $\sigma$ representing the standard deviation for each visible node. The equation for the joint energy becomes: 
\begin{align} E_{GRBM}(v,h) = \frac{\norm{v - a}^2}{2\sigma^2} - b^Th - \frac{v^T}{\sigma^2}Wh \end{align} 
where $\norm{\cdot}$ is the Euclidean norm. This equation is similar to the binary energy except that the visible states are divided by $\sigma^2$ as a form of normalization, and the first term is replaced by $\frac{\norm{v - a}^2}{2\sigma^2}$ which serves as a parabolic containment of the visible states. This means that the overall energy increases the further the visible states $v$ are from the visible biases $a$. Whereas the states in a binary RBM are bounded by 0 and 1, it is important that the visible states of a Gaussian RBM are restrained by this parabolic term in order to prevent them from trailing off.

Similar to the binary RBM, the conditional probability of the hidden states given the visible states is:
\[ p(h=1 | v) = \sigma(b + W^T\frac{v}{\sigma^2}) \]
whereas the conditional probability of the visible states given the hidden states is:
\[ p(v \ | \ h) = \mathcal{N}(a + Wh, \sigma^2) \]
where $\mathcal{N}(\mu, \sigma^2)$ is a Gaussian function with mean $\mu$ and variance $\sigma^2$. In practice this amounts to adding Gaussian noise after calculating visible states. A detailed derivation of the conditional distributions can be found in \cite{wang}.

While initially there appears to be nothing inherently Gaussian about this energy equation, the probability starts to take on the form of a Gaussian function when substituted into the Boltzmann distribution:
\begin{align*}
    p(v,h) &= \frac{1}{Z}e^{-\frac{\norm{v - a}^2}{2\sigma^2} + b^Th + \frac{v^T}{\sigma}Wh}
\end{align*} 
 If we ignore the second and third terms of the energy equation, we see that the distribution of visible states follows a Gaussian distribution centered at the visible bias $a$ with variance $\sigma^2$. Moreover, it has previously been shown using these equations that the RBM is equivalent to a mixture of Gaussian (MoG) model, with the locations of each Gaussian represented by the column vectors of the weight matrix (for a detailed derivation see \cite{melchior}). 

The energy and sampling equations are usually simplified by normalizing the data to zero mean and unit variance, and then setting $\sigma = 1$. However, one can choose to calculate the $\sigma$ of the data in advance, or alternatively one can use $\sigma$ as a separate parameter which is optimized during training. While training data need not follow a Gaussian distribution, the GRBM works best when modeling this type of distribution. Fortunately, most distributions found in nature are Gaussian due to the central limit theorem.

\paragraph{Log Likelihood Estimates}

The goal of training the RBM is to maximize the likelihood of the data under the model. By tweaking the weights and biases, the RBM can associate higher probabilities with configurations found in the training data. For practical purposes, we work with the logarithm of the likelihood which allows us to write products as the sum of logarithms. Since $\log(x)$ is a monotonically increasing function, maximizing the log-likelihood is the same as maximizing the likelihood. The log-likelihood for a training sample $x$ is given by:
\begin{align*}
    \mathcal{L}(x) &= \log p(x) \\
    &= \log \sum_{h} p(x, h) \\ 
    &= \log \sum_{h} \frac{e^{-E(x,h)}}{Z} \\ 
    &= \log \sum_{h} e^{-E(x,h)} - \log Z
\end{align*}

%\begin{align*}
%     \hat{\ell} &= \ln{P(x_1, ..., x_n \ | \ \theta)} \\
%     &= \ln{\prod^n_{i=1} P(x_i \  | \ \theta)} \\ 
%     &= \sum^n_{i=1} \ln{P(x_i \ | \ \theta)} \\ 
%     &= \sum^n_{i=1} \ln{\frac{1}{Z} \sum_h e^{-E(v,h)}} \\
%     &= \sum^n_{i=1} \ln{\sum_h e^{-E(v,h)}} - \ln{Z}
%\end{align*}
using (1) and (2). By optimizing the log likelihood of the training data, we maximize the likelihood of the training samples over all possible samples in our visible space. 

In the case where there is an entire set of training samples, we take the average log-likelihood $\hat{\ell}$ by computing the expectation of the log-likelihood over all the samples:

\[ \hat{\ell} = \bigg\langle \log \sum_{h} e^{-E(x,h)} \bigg\rangle - \log Z \]

Average log-likelihood is a rigorous way of monitoring the training and convergence of an RBM and demonstrating its modeling capacity in statistical terms. The most difficult part about estimating this likelihood is calculating the partition function $Z$, which is intractable in most cases. In this study we use importance sampling to estimate $Z$ whenever calculating log-likelihood (see \cite{krause} for more details). 

\paragraph{Training algorithm}

The gradient of $\hat{\ell}$, which is given by the derivative of $\hat{\ell}$ with respect to the model parameters $\theta$, comes out to be the difference between the data-based distribution and the distribution given by the entire model:
\[\frac{d\hat{\ell}}{d\theta} \propto \bigg \langle \frac{dE(x, h)}{d\theta} \bigg \rangle_{x} - \bigg \langle \frac{dE(v, h)}{d\theta} \bigg \rangle_v \]

where $x$ are explicit training samples and $v$ are samples from the model distribution. 

Computing the partial derivatives of the energy function with respect to each parameter provides us with the gradient approximations for the weights and biases:
\[ \frac{\partial\hat{\ell}}{\partial a} \propto \langle {x} - a \rangle_{{x}} - \langle v - a \rangle_v \]
\[ \frac{\partial\hat{\ell}}{\partial b} \propto \langle p(h=1|{x}) \rangle_{{x}} - \langle p(h=1|v) \rangle_v \]
\[ \frac{\partial\hat{\ell}}{\partial W} \propto \langle {x} \ p(h=1|{x})^T \rangle_{{x}} - \langle v \ p(h=1|{v})^T \rangle_v \]
\begin{multline*} 
\frac{\partial \hat{\ell}}{\partial \sigma} \propto \bigg \langle \frac{\norm{x-a}-2x^TW p(h=1|x)}{\sigma^3} \bigg \rangle_{{x}} \\ - \bigg \langle \frac{\norm{v-a}-2v^TW p(h=1|v)}{\sigma^3} \bigg \rangle_v 
\end{multline*}

These gradients are used as the update rules for the RBM. Typically training is done in batches, and the gradients are computed after each batch. The gradients are then added to the existing RBM parameters which gives rise to a new set of weights and biases. While the first term for each gradient can be calculated directly from the training data, the second term is almost always intractable, as it requires independent samples from an unknown model distribution. 

Several algorithms are currently available for estimating the second term. The most widely used algorithm is Contrastive Divergence. In Contrastive Divergence, sampling between visible and hidden layers is used to create a Markov chain (as shown in Figure \ref{fig:sampling}). The Markov chain is initialized at a training point, and the conditional probabilities are used to get the visible and hidden states after $k$ iterations, in a process known as Gibbs sampling. The value of these layers after sampling serves as a sufficient estimation of the model's expectation. Typically $k$ is set to 1, as it provides a good estimation of the gradient and also minimizes computation time.

Persistent Contrastive Divergence (PCD) is another algorithm which has been proposed for estimating the model distribution. Instead of restarting the Markov chain for each data point, one persistent Markov chain is retained in memory and extended after each batch. While its success has mostly been reported in binary RBMs \cite{tieleman}, some have had success in using PCD to train Gaussian RBMs \cite{courville}. A reported improvement on PCD is the Parallel Tempering algorithm (PT), which uses multiple Markov chains at different temperatures with a certain probability that states from different chains will swap \cite{cho2}. The reasoning behind PT is that having different temperatures will ensure that the Markov chain is fully exploring high energy states. Though these algorithms have all been successfully implemented in binary RBMs, there is little evidence that they work in Gaussian RBMs. In practice, we found that both PCD and PT led to divergence unless a very small learning rate was used ($\alpha = 0.0001$), which led to long training times with marginal improvements in likelihood. Overall, CD-1 was simplest to use and almost never led to divergence. Our results match those in \cite{melchior} which conclude that CD is the best algorithm currently suited for training GRBMs.

\paragraph{Monte Carlo sampling of an RBM}

Because the RBM provides us with a measure of energy, we can sample new configurations from a trained RBM using the Metropolis Monte Carlo algorithm. This is done by initializing a random configuration and evaluating its energy. Then for each iteration, a new sample $x'$ is generated by adding a random displacement vector to the previous sample $x$ and evaluating the energy of the new sample. If the energy is lower than the previous, the new sample is accepted as part of the simulation. If not, the acceptance ratio is calculated using the Boltzmann distribution:
\[\frac{p(x')}{p(x)}=\exp{\left(\frac{E(v,h)-E(v',h')}{k_BT}\right)} \]

Finally we generate a random number between 0 and 1, and if our number falls below this acceptance ratio, we keep the configuration as part of our ensemble. The Boltzmann constant $k_B$ and temperature parameter $T$ are usually set to one, though they may be useful in simulating higher temperatures or exploring higher energy states.
%In some cases, it makes more sense to use the free energy of a given configuration $F(v)$ instead of computing the hidden layer and evaluating the joint energy. In most of the cases tested, the difference between these two calculations was negligible. This is because the free energy equation uses the expected value of the hidden layer as an estimation for the sum of joint energies over all hidden states.

The benefit of this Metropolis Monte Carlo method is that the normalization for the RBM need not be known. Whereas the true probability density of the joint states requires knowledge of the partition function $Z$, which is intractable in most cases, the term cancels when calculating the acceptance ratio at each step in the simulation. The result is a Monte Carlo simulation which follows the probability density given by the model.

%Furthermore, the proposed simulation method does not require a trajectory but only a collection of random points from the target distribution, since there is no temporal element involved in the training of the RBM. Therefore, this simulation method is able to add a time component to non-temporal molecular data, turning a sampled distribution into a molecular Monte Carlo simulation. In future studies it may worthwhile to attempt to use a recurrent neural network to learn a time-based simulation, though that is outside the current scope of discussion.

\paragraph{Computational complexity}

As we can see, the energy calculations for an RBM are much simpler than the energy calculations for a quantum system. A typical Hartree-Fock calculation has computational complexity $\mathcal{O}(n^4)$, where $n$ is the total number of basis functions, since the number of two-electron integrals necessary to build the Fock matrix is $n^4$ \cite{echenique}. In practice this often becomes $\mathcal{O}(n^3)$ as most programs will ignore integrals that are close to zero. Depending on the choice of basis set and the size of the atoms, one can have anywhere from 3 to 100 basis functions for one given atom. Therefore the complexity becomes $\mathcal{O}(b^3a^3)$ where $b$ is the average number of basis functions per atom and $a$ is the number of atoms. Because the complexity scales at least cubically with system size, these calculations become quite expensive as one increases either the atoms or basis functions included.

Meanwhile, RBM energy computations have complexity $\mathcal{O}(NH)$ where $N$ is the number of visible nodes and $H$ is the number of hidden nodes. In the proposed method, only three nodes are needed per additional atom, as compared to the large number of basis functions needed in Hartree-Fock calculations. Thus the complexity is $\mathcal{O}(3aH)$, expressed in terms of the number of atoms $a$. If we take the number of hidden nodes to be the same as the number of visible nodes, then the computational complexity grows quadratically with system size, and increasing atom size has no impact on the complexity of  calculations. Moreover these calculations are performed using matrix multiplication which is simpler than integration. 

\begin{table*}[b]
\centering
\begin{adjustbox}{width=\textwidth,center=\textwidth}
\small
\def\arraystretch{1.8}
\begin{tabular}{ |c|c|c|c|c|c| } 
 \hline
 \textbf{Visible layer} & \textbf{Hidden layer} & $\mathbf{p(h|v)}$ & $\mathbf{p(v|h)}$ & $\mathbf{E(v,h)}$ & \textbf{References} \\ 
 \hline
 Binary & Binary & $\sigma(b + W^Tv)$ & $\sigma(a + Wh)$ & $-a^Tv - b^Th - v^TWh$ & \cite{salakhutdinov} \cite{hinton} \cite{larochelle}\\
 \hline
   \rule{0pt}{4.4ex} Gaussian & Binary & $\sigma(b+W^T\frac{v}{\sigma})$ & $\mathcal{N}(a + Wh, \sigma^2)$ & $\frac{\norm{v - a}^2}{2\sigma^2} - b^Th - \frac{v^T}{\sigma}Wh$ & \cite{wang} \cite{zhang} \cite{cho} \rule[-3ex]{0pt}{0pt}\\ 
 \hline 
  \rule{0pt}{4.4ex} Gaussian & Gaussian & $\mathcal{N}(b+\sigma_h W^T\frac{v}{\sigma_v}, \sigma_h^2)$ & $ \mathcal{N}(a + \sigma_v W\frac{h}{\sigma_h}, \sigma_v^2)$ &  $\frac{\norm{v - a}^2}{2\sigma_v^2} + \frac{\norm{h - b}^2}{2\sigma_h^2} - \frac{v^T}{\sigma_v}W\frac{h}{\sigma_h} $ & \cite{karakida} \cite{ogawa} \rule[-3ex]{0pt}{0pt} \\ 
 \hline
 Gaussian & ReLU & $max(0,\eta + \mathcal{N}(0, \sigma(\eta))$ & $\mathcal{N}(a + Wh, \sigma^2)$ & N/A & \cite{nair} \\
 \hline
   \rule{0pt}{8ex} Gaussian & Leaky ReLU & \begin{tabular}{@{}c@{}} $max(\mathcal{N}(c\eta, c), \mathcal{N}(\eta,1))$, \\ where $c\in (0,1)$ \end{tabular} & $\mathcal{N}(a + Wh, \sigma^2)$ & 
 \begin{tabular}{@{}c@{}}$\frac{\norm{v - a}^2}{2\sigma^2} - b^Th - \frac{v^T}{\sigma}Wh$ \\ $+ \sum_{h_j > 0}(\frac{h_j^2}{2}+\log \sqrt{2\pi}) + \sum_{h_j \leq 0}(\frac{h_j^2}{2c}+\log \sqrt{2c\pi})$ \end{tabular} & \cite{li} \rule[-7ex]{0pt}{0pt}\\
 \hline
 \rule{0pt}{8ex} Gaussian & (Noisy) Tanh & $\mathcal{N}(\tanh(\eta),1)$ & $\mathcal{N}(a + Wh, \sigma^2)$ & \begin{tabular}{@{}c@{}}$\frac{\norm{v - a}^2}{2\sigma^2} - b^Th - \frac{v^T}{\sigma}Wh$ \\ $+ \sum^J_{j=1} (h_j\tanh^{-1}(h_j) + \frac{1}{2}\ln(1-h_j^2))$ \end{tabular} & See appendix. \rule[-7ex]{0pt}{0pt} \\ 
 \hline
\end{tabular}
\end{adjustbox}
\caption{Different versions of RBMs found in the literature, where $\eta = b+W^T\frac{v}{\sigma}$. This table is intended to represent four main activation functions which are prominent in neural networks: linear (Gaussian), sigmoid (binary), ReLU, and Tanh. In the design of a restricted Boltzmann machine, one typically starts with a function that defines the overall energy $E(v,h)$, from which the conditional probabilities are derived. Advantages of different activation functions}
\label{table:methods}
\end{table*}

\begin{figure*}[b]
    \centering
    \fbox{\includegraphics[scale=0.42]{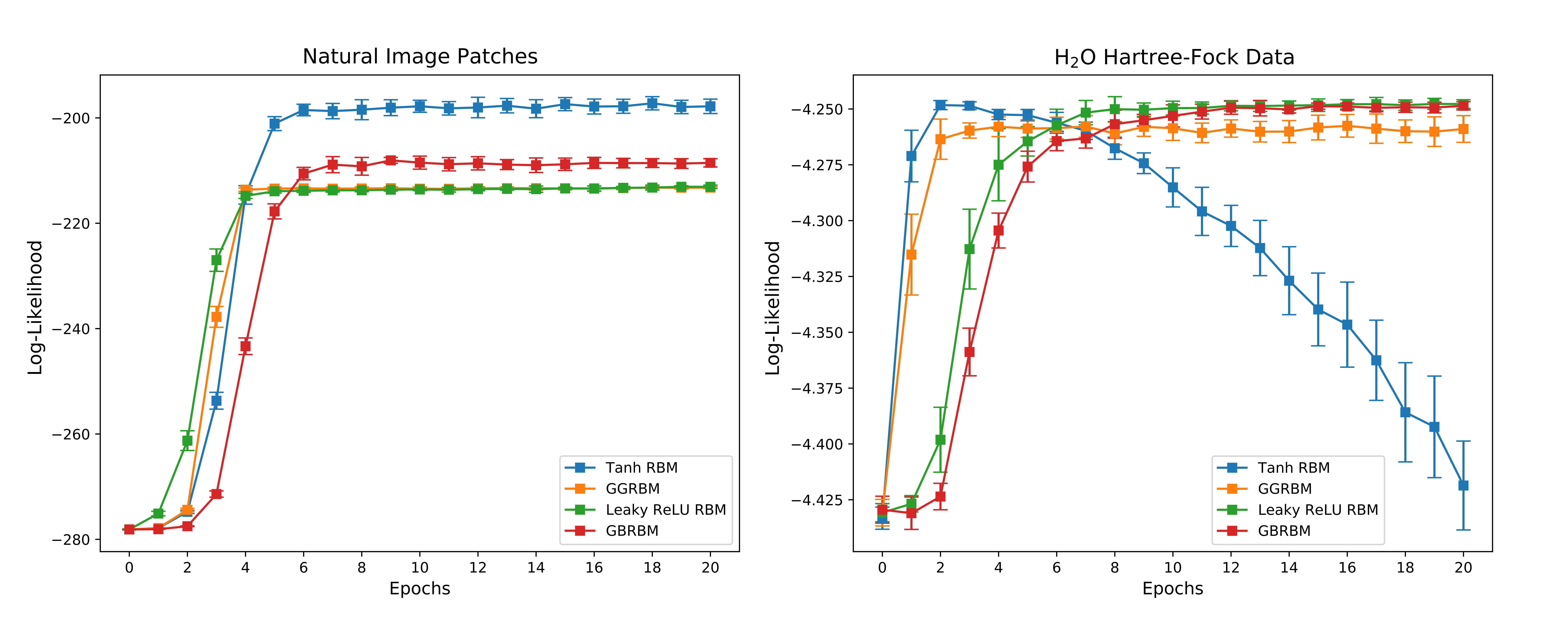}}
    \caption{Comparison of RBM Performance on Natural Image Patches (left) and H$_2$O Hartree-Fock Data (right). Learning on natural image patches was performed using 196-196-RBMs, which H$_2$O learning was performed using 3-8-RBMs. In both cases, Tanh RBM achieved the maximum overall log-likelihood, however it has trouble maintaining convergence on datasets with few parameters. Though the GRBM had only the second best performance on both datasets, it trains slower and shows better consistency across different data.}
    \label{fig:rbm_LL}
\end{figure*}

\begin{figure*}[b]
    \centering
    \fbox{\includegraphics[scale=0.51]{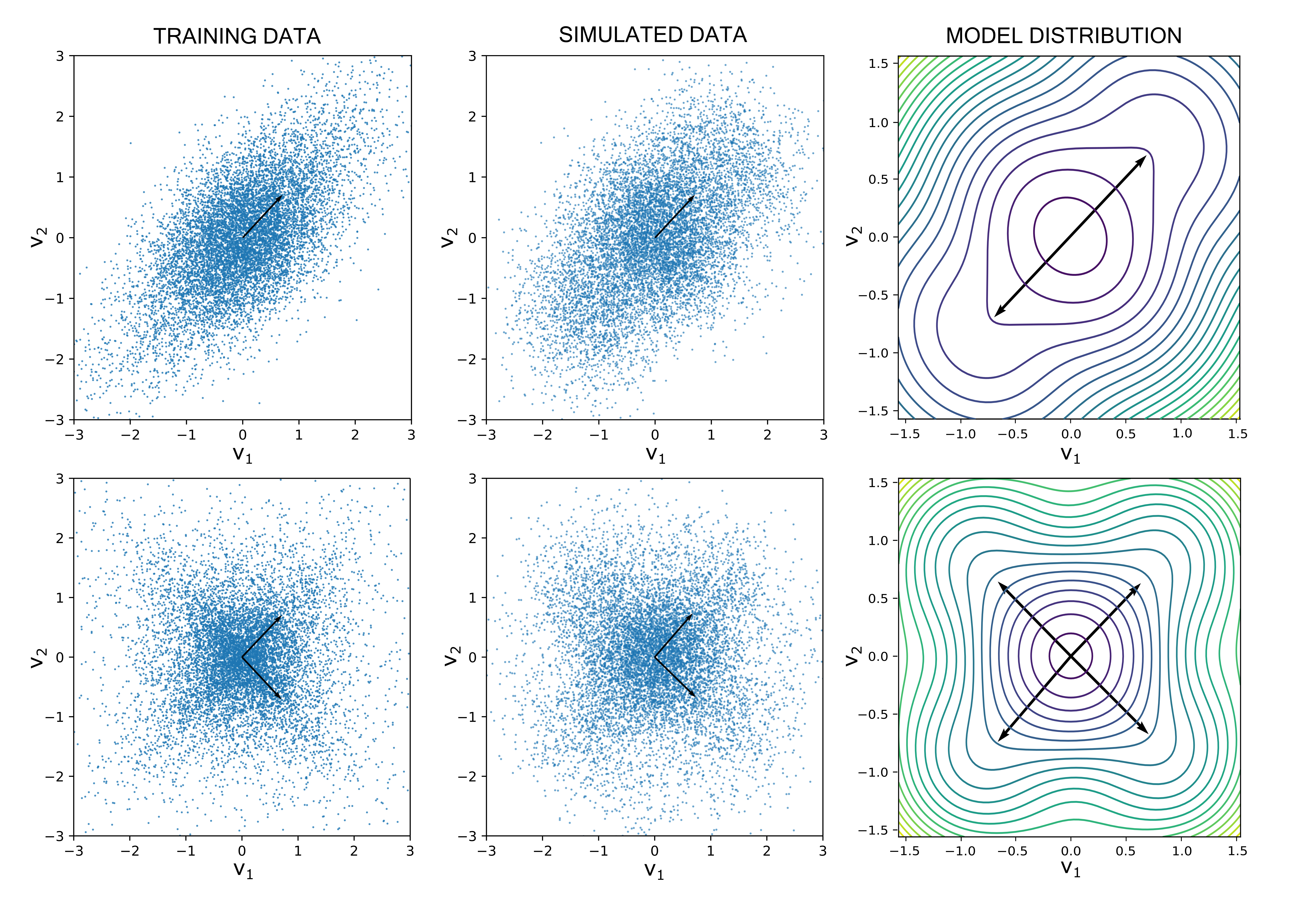}}
    \caption{Training and simulation of a GRBM using two different toy sets of data. The top row was trained using a GRBM-2-2 (two visible nodes and two hidden nodes), while the bottom row was trained using a GRBM-2-4 (two visible nodes and four hidden nodes). Each hidden node represents an independent component in the data, shown by the arrows on the model distributions in the third column. The independent components of the simulated data (shown by the arrows on the second column) align with the independent components on the training data (first column). The RBM parameters were averaged over 100 trials of training.}
    \label{fig:components}
\end{figure*}

\begin{figure*}[b]
    \centering
    \fbox{\includegraphics[scale=0.39]{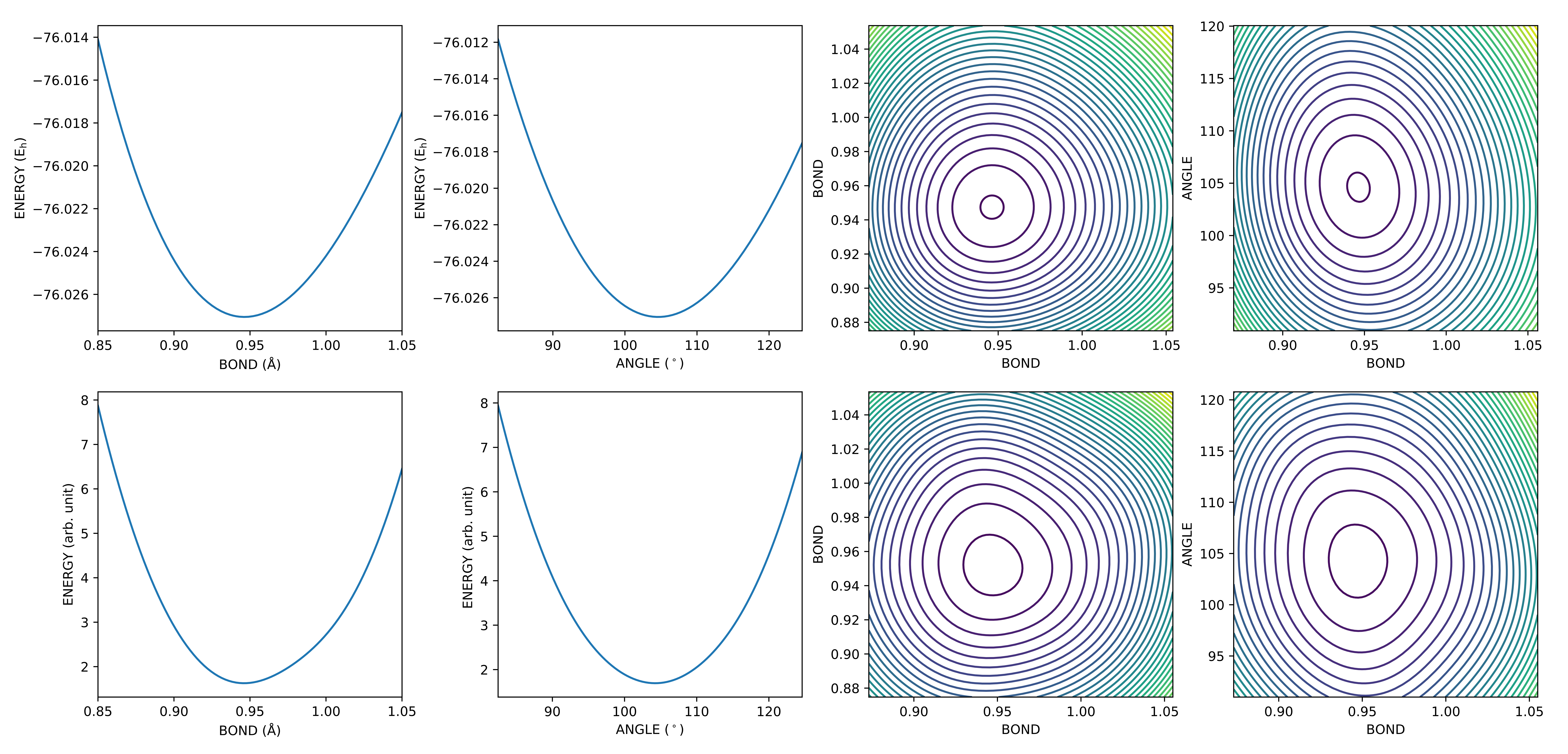}}
    \caption{Bond and angle energies evaluated using Hartree-Fock (top row) and a GRBM-3-6 (bottom row). The GRBM is able to provide a close approximation of the original energy contour based on a limited set of points sampled from the original distribution.}
    \label{fig:hartreefock}
\end{figure*}

\begin{figure*}[t]
    \centering
    \fbox{\includegraphics[scale=0.34]{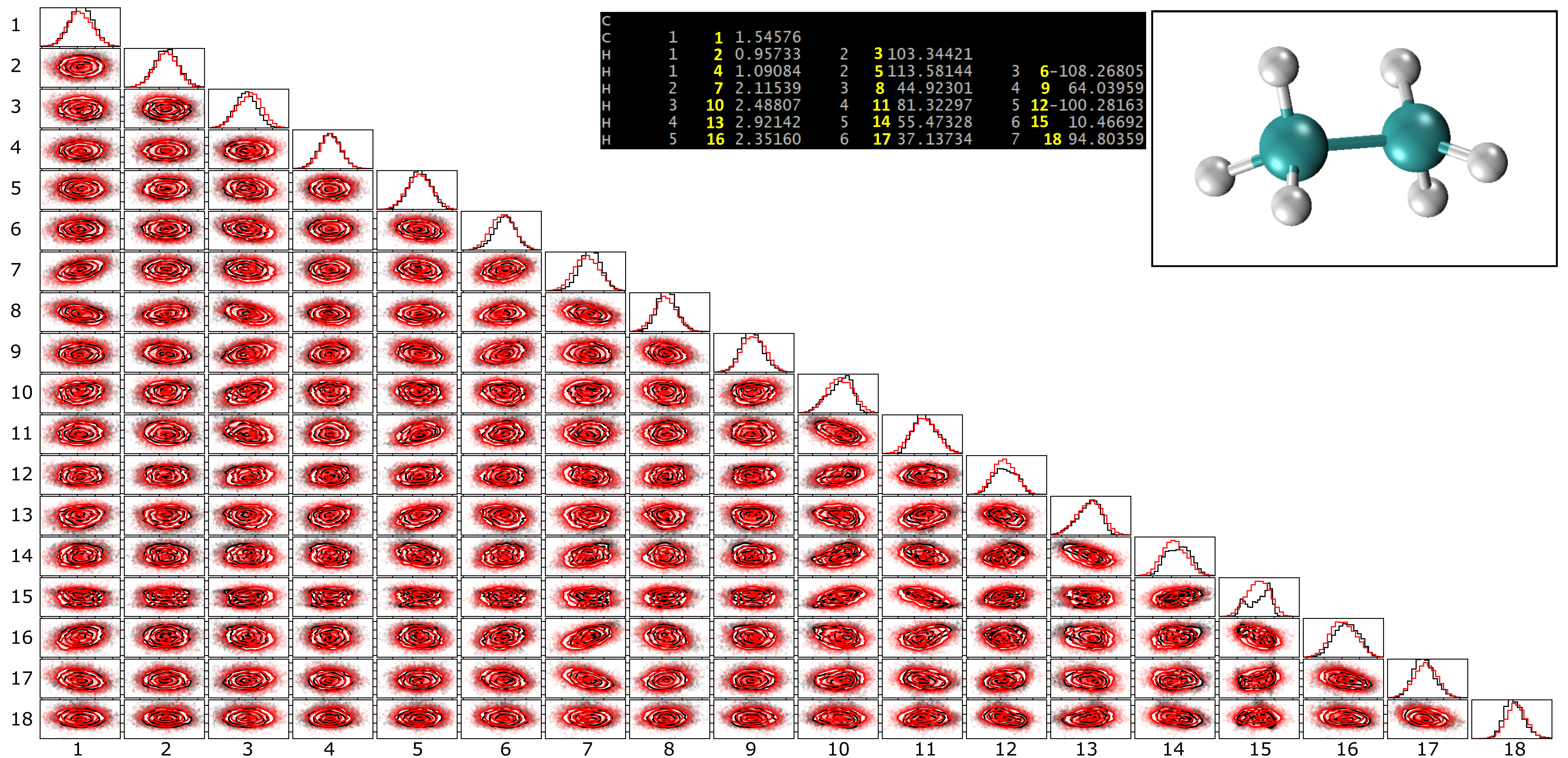}}
    \caption{Ethane geometries generated by Hartree-Fock and GRBM. An ethane molecule has 18 internal coordinates, represented by the entries in the z-matrix above. That specific set of coordinates results in the ethane molecule drawn in the top right corner. This figure displays the posterior distributions for each pair of parameters for the ethane molecule generated by Hartree-Fock (black) and GRBM (red). The GRBM is able to reproduce the molecular geometries by fitting Gaussians to the original data.}
    \label{fig:ethane}
\end{figure*}

\paragraph{Generating training datasets} 
To assess the validity of the methodology and accuracy of training over a wide range of distributions, we trained the RBM on a set of toy datasets. In the first toy dataset, random samples were taken from a Laplacian distribution and fitted to the line $y = x$, with Gaussian noise of unit variance added in both dimensions. The second dataset was generated by sampling from two independent Laplacian distributions and using a mixing matrix to merge the sources. An observable $x$ is generated from independent Laplacian sources $s = (s_1,s_2)$ using a mixing matrix $A$:
\[x = As \]
where $A$ is a linear transformation from the source space to the observable space. Given a set of observables, the RBM is capable of estimating the mixing matrix and recovering source data \cite{melchior}. In the context of this task, independent component analysis is the standard computational method for separating observed data into source components. For this reason, the independent components of the original data were compared with the independent components of the data generated by RBM. Independent component analysis was performed using the FastICA algorithm. % To  assess the validity of the methodology and accuracy of training over a wide range of distributions, several sample distributions were generated via Monte Carlo, where an energy equation was supplied as a function of the other parameters, whether they symbolize physical measurements or quantum states. 

Moving to more complex data, different RBMs were also evaluated on natural image patches taken from \cite{melchior}. These image patches come from randomly sampling the greyscale images found in the van Hateran natural image database \cite{vanhateran}. Each patch consists of 14 x 14 pixels, and for our purposes we used only 10,000 of these patches during training.

Meanwhile, molecular training data for \texorpdfstring{H\textsubscript{2}O}{H2O} and ethane was generated by performing molecular Monte Carlo simulations and evaluating the energy of each proposed sample using the restricted Hartree-Fock method. The Hartree-Fock calculations made use of the basis set cc-pVDZ because it had the best agreement with literature values for the bond angle and bond length of \texorpdfstring{H\textsubscript{2}O}{H2O}. These calculations were run using the PySCF module for quantum chemistry.

\paragraph{Different RBMs}
There are many types of RBMs apart from those already discussed. By altering the equation for the overall energy, a variety of hidden and visible units can be implemented. Table \ref{table:methods} summarizes the energy equations and conditional probabilities for a few different types of RBMs. These RBMs are mostly defined by the activation functions used to calculate the hidden layer, including sigmoid, Gaussian, ReLU.

Noticing the absence of an RBM with Tanh activation function, we derived a new energy function and conditional probabilities for an RBM with Tanh hidden units using the method outlined in \cite{ravanbaksh} (see appendix for derivation). After developing the Tanh RBM, we implemented the full range of RBMs defined in Table \ref{table:methods}, which use common activation functions from machine learning. This implementation can be found at \url{https://github.com/peter1255/RBM_chem}. 

Depending on the correlations found within the data, different activation functions may be better for modeling different data. It has been shown that non-linear hidden units expand the capabilities of the RBM by allowing the model to represent non-linear correlations between visible units \cite{ravanbaksh}. The type of non-linearity chosen may have an effect on the model's representation of the data, though a more detailed comparison of RBMs is needed. 

%------------------------------------------------

\section{Results and Discussion}

\paragraph{Comparison of RBMs}

The RBMs found in Table \ref{table:methods} were implemented and trained on both natural image data and H$_2$O geometries. During 20 epochs of training, the performances of RBMs were compared through thorough calculations of the log-likelihood. These results are shown in Figure \ref{fig:rbm_LL}.  

First proposed by \cite{nair}, the ReLU RBM trains well but lacks an appropriate joint energy equation \cite{su}, making it impossible to calculate log-likelihood or generate simulations. Though we tried using the Gaussian joint energy equation (3) in test simulations not shown here, values quickly diverged as large hidden terms outweighed the parabolic containment term. Making a slight modification of ReLU into Leaky ReLU gives a viable energy equation which can be successfully trained and used for both sampling and calculating log-likelihood \cite{li}. While this version of the RBM has not yet reached widespread use, we show in \ref{fig:rbm_LL} that it performs on a level comparable with the previously established GRBM. 

Out of all the RBMs tested, the Tanh RBM achieved the best maximum log-likelihood on both natural image data and H$_2$O molecular geometries. However, when trained on data sets with few parameters, like H$_2$O data, the Tanh RBM has trouble maintaining convergence and tends to diverge after reaching its maximum. While the Tanh RBM shows great promise for representing a variety of data, additional research must be done to learn how to prevent divergence in general cases. 

The model that achieves second-best performance in terms of log-likelihood is the Gaussian RBM, which is quite close in the case of molecular data (Figure \ref{fig:rbm_LL}).  For this reason, we selected this version of the model to test the proposed simulation method on molecular data.

%Training of the RBM on each data set was performed using the Contrastive Divergence algorithm (CD). In all cases, this was done for only one iteration of Gibbs sampling (CD-1). The accuracy of the training was evaluated by comparing the input to the visible layer after CD-1. With the GRBM, 10 epochs was typically sufficient to reach maximum training accuracym shown in \ref{table:}, while avoidng overfitting. ($\sim0.60$ MSE normalized). While much closer training accuracies were possible, they usually led to overfitted simulations.  %\# of hidden neurons 
%\# of samples , batch size etc

%delta simulation, acceptance rate etc

%The number of hidden nodes sufficient for training is usually a little more than double the number of visible nodes. 

\paragraph{Modeling two-dimensional mixtures}
Figure \ref{fig:components} compares toy datasets with the datasets generated by a trained GRBM using the Monte Carlo method outlined above. The left column displays the original datasets, while the middle column displays the data generated by RBM. The two datasets appear quite similar to one another, though some of the pointedness in the original data is lost through the RBM's Gaussian probability distribution. After performing independent component analysis on both training and simulated data, we can see that the independent components (shown using arrows) from the simulated data (middle column) match those from the original data (left column).

Then, evaluating the energy of each point in the visible space using equation (3), we graphed the energy contour represented by the GRBM (right column). By moving along this energy contour we are able to generate the states represented in the middle column. Because the Monte Carlo method favors lower energy states, most of the samples lie within the middle of the contour. Furthermore, the arrows represented by each hidden node of the RBM match the independent components of the reconstructed data, showing that the Monte Carlo sampling method preserves the original components of the data.

Furthermore, very few hidden nodes were required for reproducing the data. In the case of the linear distribution which has only one independent component, only two hidden nodes were needed. In the case of the cross shaped distribution, only four hidden units were needed. In general, two hidden nodes are needed for each independent component in the data: because the hidden nodes are binary, they can only represent their positive vector span. Therefore an additional hidden node is needed to represent the opposite direction. This also provides a useful heuristic for determining the number of hidden nodes to include in an RBM. In this case, using any more arrows would be redundant, as the arrows begin to overlap. 

%The training data was then expanded into three dimensions in order to add another parameter of learning for the RBM. Figure 3 represents a hypothetical molecular dynamics problem for which an RBM may be useful. When H2O is modeled after a coupled oscillator, there are three normal modes that correspond to internal movements of the water molecule: symmetric stretching, antisymmetric stretching, and bending. If these are taken to be the three hidden sources of data, the mixing matrix estimated by the RBM can tell us how these modes are combined.

\paragraph{Modeling molecular geometries}
After learning toy distributions, the RBM was trained on molecular geometries of H\textsubscript{2}O generated by Hartree-Fock. Since the geometry of H\textsubscript{2}O is defined by three parameters (two bonds and one angle), three visible nodes were included in the GRBM. The GRBM was trained on a set of 10000 geometries. Figure 3 displays the overall bond and angle energies evaluated using Hartree-Fock (top row) and GRBM (bottom row). 

Not only is the energy contour of the RBM similar to the energy determined by quantum methods, the RBM was able to pick up on subtleties including the rotated parabola which defines the bond and angle energies (first and second columns). Looking at the bond and angle energies given by the RBM, the left arm of the parabola shows the energy associated with atom-atom repulsion, while the right side of the graph shows the energy associated with atom-atom attraction. Both methods match the literature values for the bond length and angle of H$_2$O.

Similar to the previous figure, energy contours (third and fourth columns) were generating by fixing one parameter to its given minimum and evaluating the energy with respect to the other two parameters. As we can see, the energy contours inferred by GRBM are quite similar to the original contours given by Hartree-Fock. Though the stratification of the contour layers is slightly different (a fundamental limitation of the RBM model structure) the overall shape of the contours is quite similar. The GRBM accurately represents the relationship between different data parameters.  

Finally, we use the RBM to model a larger molecule: ethane. Though it only consists of 6 atoms, ethane requires 18 internal coordinates to represent all its various bonds, angles, and dihedrals. In general, a molecule with $n$ atoms requires $3n-6$ unique coordinates, which can be stored in the form of a z-matrix. Therefore we train an RBM with 18 visible nodes on the internal ethane coordinates, and then sample from it to generate new geometries. 

Figure \ref{fig:ethane} shows the difference between the original dataset (black) and the GRBM generated data (red). The reconstructed data is able to approximate molecular geometries by fitting Gaussian curves to the original data. For simple unimodal data, the GRBM does a good job of reproducing the existing shape of the data. However, for more complex distributions (i.e. the bimodal dihedral angle found in coordinate 15 of \ref{fig:ethane}) the GRBM is unable to provide a fit for this complex pattern and instead regresses to the mean in between the two modes. Nevertheless, the rest of the marginal distributions are well represented, and the GRBM captures the essence of the joint distributions. Through our Monte Carlo sampling algorithm, the GRBM provides an accurate picture of the dynamics of an ethane molecule. 

Using the RBM it is easy to evaluate the energies of any configurations at a fast speed. By scaling the energy units of the RBM to proper Hartree units, we could use the RBM to represent bond and angle energies without having to perform any quantum calculations. Moreover, because of the low computational cost in generating states of a water molecule, the technique here could also be extended to model an entire body of liquid water in an aqueous solution. 

%pausing training at different moments 

%Ideas:
%- subsampling same distribution (what happens if you use less or you just remove an important chunk of the graph?)
%- exploration of hidden filters? plot showing how often each hidden unit is active as well as what values they correspond to
%- fraction of confs on which a hidden unit is active vs. fraction of hidden units (see nair and hinton)
%- simulated annealing to find the ground state / equilibrium state... training an RBM then sampling at high temperatures and gradually lowering temperature
%- further investigate relationship between energy function and wavefunction
%- binary rbm qubits / electron spins 
%- make a parity plot

%------------------------------------------------

\section{Conclusion}
In this paper, we have demonstrated the usefulness of RBMs in modeling complex molecular systems. Because the model is adaptable to a wide variety of data, the proposed methodology can be used for a variety of problems in chemistry. Furthermore, the relative simplicity and efficiency of the model should make it accessible to a wider scientific audience. The RBM is a useful method for generating a complete molecular ensemble given a sparse set of data. We hope that the RBM will allow for a new cycle of experiment and theory, where samples generated from experiments are treated with RBMs to get more information about ensemble systems. 

\paragraph{Note} The software package used for this study along with a tutorial for how to use it to model molecular data is available on Github at \url{https://github.com/peter1255/RBM_chem}.

%--------------------------------------------------------------------------
%	REFERENCE LIST
%--------------------------------------------------------------------------

%---------------------------------------------------------------------------

\onecolumn

\section{Appendix}

\paragraph{Derivation of Tanh energy} First we define the activation function $f(\eta_j) = \tanh(\eta_j)$ where $\eta_j = b_j + \sum_{i=1}^I W_{ij}v_j$ assuming data is normalized to unit variance. The inverse function is $f^{-1}(h_j)=\tanh^{-1}(h_j)$. The corresponding antiderivatives are: 
\begin{align*}
F(\eta_j) &= \int f(\eta_j)d\eta \\
&= \ln{\cosh{\eta_j}} + C \\
F^*(h_j) &= \int f^{-1}(h_j)dh \\
&= h_j\tanh^{-1}(h_j) + \frac{1}{2}\ln(1-h_j^2) + C      
\end{align*}

Meanwhile, the activation for the visible units is linear so that $\overline{f}(\nu_i)=\nu_i$ where $\nu_i =  a_i + \sum^J_{j=1}W_{ij}h_j$. Following the same step we get that $\overline{F}(\nu_i) = \frac{1}{2}\nu_i^2$ and $\overline{F}^*(v_i) = \frac{1}{2}v_i^2$.

From Ravanbakhsh et al. \cite{ravanbaksh}, the conditional distribution is defined as:
\[ p(h_j|\eta_j) = \exp(h_j\eta_j - F(\eta_j) - F^*(h_j) + g(h_j)) \]

Replacing the values of $F(\eta_j)$ and $F^*(h_j)$ we get the conditional probability: 
\begin{align*}
p(h_j|\eta_j) &= \exp(h_j\eta_j - \ln{\cosh{\eta_j}} \\ -& h_j\tanh^{-1}(h_j) 
- \frac{1}{2}\ln(1-h_j^2) + g(h_j)) \\
&= \exp(\tanh(\eta_j)\eta_j - \ln{\cosh{\eta_j}} \\ -& \tanh(\eta_j)\tanh^{-1}(\tanh(\eta_j)+g(h_j)) 
- \frac{1}{2}\ln(1-h_j^2) + g(h_j)) \\
&= \exp(\tanh(\eta_j)\eta_j - \ln{\cosh{\eta_j}} \\ -& \tanh(\eta_j)\tanh^{-1}(\tanh(\eta_j)) 
- \frac{1}{2}\ln(1-h_j^2) + g(h_j)) \\
&= \exp(- \ln{\cosh{\eta_j}} - \frac{1}{2}\ln(1-h_j^2) + g(h_j)) \\
&=\frac{1}{\cosh{\eta_j}}\frac{1}{\sqrt{1-\tanh^2{\eta_j}}}\exp(g(h_j)) \\
&=\frac{1}{\cosh{\eta_j}\sech{\eta_j}}\exp(g(h_j)) \\ 
&= \exp(g(h_j))
\end{align*}

For simplicity we define $g(h_j) = -\frac{1}{2}h_j^2 - \log \sqrt{2\pi}$ so that the conditional probability distribution becomes a Gaussian with mean $h$ and unit variance, meaning that Gaussian noise is added to the hidden units after applying the Tanh activation. Similarly we get that $p(v|\nu)=\mathcal{N}(\nu, 1)$, as expected for the Gaussian visible layer.

The joint energy equation was generalized in Yang et al.\cite{yang} through the following: 
\[E(v,h) = -v^TWh + \overline{F}^*(v) + F^*(h) \]

Replacing $\overline{F}^*$ and $F^*$ we get the energy for a joint configuration of the tanh RBM: 

\[E(v,h) = -v^TWh + \frac{1}{2}||v||^2 + \sum^J_{j=1} (h_j\tanh^{-1}(h_j) + \frac{1}{2}\ln(1-h_j^2)) \]

From the joint energy we can derive the marginal probability of a given visible configuration $v$: 

\begin{align*}
p(v) &= \frac{1}{Z} \int_h e^{-E(v,h)}dh  \\
&= \frac{1}{Z} \int_h e^{v^TWh - \frac{1}{2}||v||^2 - \sum^J_{j=1} (h_j\tanh^{-1}(h_j) - \frac{1}{2}\ln(1-h_j^2))}dh \\
&= \frac{1}{Z} e^{-\frac{1}{2}||v||^2} \int_h e^{v^TWh - \sum^J_{j=1} (h_j\tanh^{-1}(h_j) + \frac{1}{2}\ln(1-h_j^2))}dh \\
&= \frac{1}{Z} e^{-\frac{1}{2}||v||^2} \int_h e^{\sum^J_{j=1} \eta_jh_j - h_j\tanh^{-1}(h_j) - \frac{1}{2}\ln(1-h_j^2)}dh \\ &= \frac{1}{Z} e^{-\frac{1}{2}||v||^2} \int_h \prod^J_j e^{\eta_jh_j - h_j\tanh^{-1}(h_j) - \frac{1}{2}\ln(1-h_j^2)}dh \\
&= \frac{1}{Z} e^{-\frac{1}{2}||v||^2} \prod^J_j \int_{h_j} e^{\eta_jh_j - h_j\tanh^{-1}(h_j) - \frac{1}{2}\ln(1-h_j^2)}dh_j \\
&= \frac{1}{Z} e^{-\frac{1}{2}||v||^2} \prod^J_j e^{\eta_j\tanh(\eta_j) - \tanh(\eta_j)\tanh^{-1}(\tanh(\eta_j)) - \frac{1}{2}\ln(1-\tanh^2(\eta_j))} \\
&= \frac{1}{Z} e^{-\frac{1}{2}||v||^2} \prod^J_j e^{ - \frac{1}{2}\ln(1-\tanh^2(\eta_j))} \\
&= \frac{1}{Z} e^{-\frac{1}{2}||v||^2} \prod^J_j e^{ - \frac{1}{2}\ln(1-\tanh^2(b_j + \sum_{i=1}^I W_{ij}v_j))} \\
\end{align*} 

where Z is the partition function. 

Since the marginal energy is given by \[E(v) = -\log p(v) = F(v) - \log(Z)\] where $F(v)$ is the free energy of the given configuration, we have the following expression for the free energy

\[F(v) = \frac{||v||^2}{2} + \sum^J_j \frac{1}{2}\log(1-\tanh^2(b_j + \sum_{i=1}^I W_{ij}v_j))\]


\begin{thebibliography}{99} 

\bibitem{cho}
    Cho, K.H., Raiko, T., Ilin, A. Gaussian-Bernoulli deep Boltzmann machine. Proceeding of the The International Joint Conference on Neural Networks (IJCNN) 1–7 (2013).


\bibitem{cho2}
    Cho, K., Raiko, T., \& Ilin, A. Parallel tempering is efficient for learning restricted Boltzmann machines.  Proceedings of the International Joint Conference on Neural Networks (IJCNN) 3246-3253 (2010).

\bibitem{courville}
    Courville, A., Bergstra, J., \& Bengio, Y. A spike and slab restricted Boltzmann machine. Proceeding of the Society for Artificial Intelligence and Statistics (2011).
    
\bibitem{echenique}
    Echenique, P. \& Alonso, J. L. A mathematical and computational review of
    Hartree-Fock SCF methods in quantum chemistry. Molecular Physics 105:3057-3098 (2007). 

\bibitem{hinton}
    Hinton, G. E. \& Salakhutdinov, R. R. Reducing the Dimensionality of Data with Neural Networks. Science 313;5786: 504–507 (2006).    
    
\bibitem{hinton2}
    Hinton, G.E. Scholarpedia, 2(5):1668 (2007).

\bibitem{karakida}
    Karakida, R., Okada, M., Amari, S. Dynamical analysis of contrastive divergence learning: Restricted Boltzmann machines with Gaussian visible units. Neural Netw. 79:78-87 (2016).
    
\bibitem{krause}
    Krause, O., Fischer, A., \& Igel, C. Algorithms for estimating the partition function of restricted Boltzmann machines. Artifical Intelligence 278 (2020). 

\bibitem{larochelle}
    Larochelle, H.; Bengio, Y. Classification using discriminative restricted Boltzmann machines. Proceedings of the 25th international conference on Machine learning - ICML:536 (2008).

\bibitem{li}
    Li, C.L., Ravanbakhsh, S., \& Poczos, B. Annealing Gaussian into ReLU: a new sampling strategy for leaky-ReLU RBM. arXiv preprint arXiv:1611.03879 (2016).

\bibitem{melchior}
     Melchior, J., Wang, N., Wiskott, L. Gaussian-binary restricted Boltzmann machines for modeling natural image statistics. PLOS ONE 12;3 (2017).

\bibitem{nair}
    Nair, V. \& Hinton, G. E. Rectified linear units improve restricted boltzmann machines. In ICML (2010).    

\bibitem{ogawa}
     Ogawa, S., \& Mori, H., A Gaussian-Gaussian-restricted-Boltzmann-machine-based deep neural network technique for photovoltaic system generation forecasting, IFAC-PapersOnLine, 52;4:87-92 (2019).

\bibitem{ravanbaksh}
    Ravanbakhsh, S. et al. Stochastic neural networks with monotonic activation functions. Proceedings of the 19th International Conference on Artificial Intelligence and Statistics (2016).

\bibitem{salakhutdinov}
    Salakhutdinov, R., Mnih, A., \& Hinton, G.E. Restricted Boltzmann machines for collaborative filtering. Proceedings of the 24th International Conference on Machine Learning (2007).

\bibitem{smolensky}
    Smolensky, P. Information processing in dynamical systems: Foundations of harmony theory. Parallel Distributed Processing, 1:194-281 (1986).

\bibitem{su}
     Su, Q. et al. Unsupervised Learning with Truncated Gaussian Graphical Models. Proceedings of the Thirty-First AAAI Conference on Artificial Intelligence (2017).

\bibitem{tieleman}
    Tieleman, T. Training restricted boltzmann machines using approximations to the likelihood gradient. In ICML 25:1064–1071 (2008).
     
\bibitem{vanhateran}
    van Hateren, J. H., \& van der Schaaf, A. Independent Component Filters of Natural Images Compared with Simple Cells in Primary Visual Cortex. In Proceedings of Biological Sciences 359-366 (1998).

\bibitem{wang}
    Wang, N., Melchior, J., \& Wiskott, L. Gaussian-binary Restricted Boltzmann Machines on Modeling Natural Image Statistics. CoRR. (2014).

\bibitem{xia}
    Xia, R. \& Kais, S. Quantum machine learning for electronic structure calculations. Nature Communications 9:4195 (2018).
    
\bibitem{yang}
    Yang, E., Ravikumar, P., Allen, G.I., and Liu, Z. Graphical models via generalized linear models. In NIPS (2012).

\bibitem{yu}
     Yu, W. et al. Generating the conformational properties of a polymer by the restricted Boltzmann machine. J. Chem. Phys. 151, 031101 (2019).
     
\bibitem{zhang}
    Ji Zhang, Hongjun Wang, Jielei Chu, Shudong Huang, Tianrui Li, Qigang Zhao, Improved Gaussian–Bernoulli restricted Boltzmann machine for learning discriminative representations. Knowledge-Based Systems 185 (2019).
 
\end{thebibliography}
\end{document}